\documentclass{amsart}

\usepackage[pagewise]{lineno}\linenumbers
\usepackage[utf8]{inputenc}
\usepackage{amsthm,amsmath,amsfonts}
\usepackage{soul,color}
\usepackage{longtable}
\usepackage{xcolor}
\usepackage{tabularx}
\usepackage{longtable}
\usepackage{booktabs}
\usepackage{mathtools}
\usepackage{mathabx} 
\newcommand{\Fq}{\mathbb{F}_q}
\newcommand{\Z}{\mathbb{Z}_4}
\newtheorem{theorem}{Theorem}[section]
\newtheorem{corollary}{Corollary}[theorem]
\newtheorem{lemma}[theorem]{Lemma}

\usepackage{color,soul,xcolor}
\usepackage{multicol}
\usepackage{enumitem}
\usepackage{hyperref}
\usepackage{caption}
\usepackage[ruled,vlined]{algorithm2e}
\usepackage{makecell}

\usepackage{comment}
\usepackage{diagbox}
\usepackage[justification=centering]{caption}
\theoremstyle{definition}

\title{An Updated Database of $\Z$ Codes }

\author{Nuh Aydin, aydinn@kenyon.edu \\
    Yiang Lu, lu1@kenyon.edu\\ 
  Vishad Raj Onta, onta1@kenyon.edu }

\date{}

\begin{document}
\nolinenumbers

\maketitle
\begin{abstract}
Research on codes over finite rings has intensified since the discovery in 1994 of the fact that some best binary non-linear codes can be obtained as images of $\Z$-linear codes \cite{Z4main}. Codes over many different finite rings has been a subject of much research in coding theory after this discovery. Many of these rings are extensions of $\Z$. As a result, an online database of $\Z$ was created in 2008 \cite{Z4Database}. The URL of the original database on $\Z$ codes has recently changed. The purpose of this paper is to introduce the new, updated database of $\Z$ codes. We have made major updates to the database by adding 8701 new linear codes over $\Z$. These codes have been found through exhaustive computer searches on cyclic codes and by an implementation of the ASR search algorithm that has been remarkably fruitful to obtain new linear codes from the class of quasi-cyclic (QC) and quasi-twisted (QT) codes over finite fields. We made modifications to the ASR algorithm to make it work over $\Z$. The initial database contained few codes that were not free. We have added a large number of non-free codes. In fact, of the 8701 codes we have added, 7631 of them are non-free. 


\end{abstract}


\section{Introduction and Motivation}

Codes over finite rings have received much attention from coding theory researchers in the last few decades. Various kinds of finite rings have been considered as the code alphabet in this research. However, the ring $\Z$ and its extensions have a special places among finite rings in coding theory.

There are several online databases for various types of codes with data on best known codes. For example, M. Grassl maintains the database at \cite{database} that holds records of best known linear codes over small finite fields of order up to 9. A  similar database is available in the Magma software \cite{magma}. Grassl also maintains a database of best known additive quantum codes over the binary field. Chen maintains a table of best known QC and QT codes at \cite{qcdatabase}. A table of nonlinear binary codes is available at \cite{NonlinearBinaryTable} There are several other databases of interest for coding theorists that are listed at \cite{database}. 

Due to the special place of $\Z$ in coding theory, a database of $\Z$ codes was created in 2008 and the paper that introduced it was published in 2009 \cite{Z4Database}. The database was populated with codes obtained from various search algorithms but it was not complete, especially in terms of non-free codes. Researchers have found many new codes over the years that have been added to the database. However, it is still incomplete and there is room for additional entries. Also, due to technical issues the server and the URL for the original database have recently changed. The new URL for the database is http://quantumcodes.info/Z4/ (We also created a database of quantum codes recently, available at http://quantumcodes.info/)

This work aims to rectify the issues summarized above by carrying out exhaustive searches for both free and non-free cyclic codes over $\Z$, as well as adapt the highly effective ASR search algorithm for $\Z$. Carrying out these searches resulted in the discovery of many new linear codes over $\Z$, as well as rediscovering several good best known linear codes through the  Gray map. Two non-linear binary codes with new parameters have been found as well, with the caveat that there are best known linear codes with equivalent parameters.

\section{Basic Definitions}
A code $C$ of length $n$ over $\Z$ is a subset of $\Z^{n}$. If $C$ is an additive subgroup of $\Z^{n}$, then C is a also submodule of $\Z$ and it is a linear code over $\Z$. A codeword is an element of $C$, and a generator matrix is a matrix whose rows generate $C$. Any linear code $C$ over $\Z$ is equivalent to a code over that ring with a generator matrix $G$ of the form
\begin{align*}
    \begin{bmatrix}
    I_{k_1} & A & B \\
    0 & 2I_{k_2} & 2C
    \end{bmatrix}
\end{align*}
where $A$ and $C$ are $\mathbb{Z}_2$-matrices and $B$ is a $\Z$-matrix. We then say that $C$ is of type $4^{k_1}2^{k_2}$, which is also the size of $C$. We express the parameters of $C$ as $[n,k_1,k_2,d_L]$ or just $[n,k_1,k_2,d]$, where $n$ is the length of $C$ and $d_L$ or $d$ is the minimum Lee distance of $C$. A $\Z$-linear code is not
necessarily a free module, a module with a basis. It is so if and only
if $k_2 = 0$. If $k_2$  is zero, we call $C$ a free code. Otherwise,  it is non-free.

The Gray map $\phi: \Z^n \rightarrow \mathbb{Z}_2^{2n}$ is the coordinate-wise extension of the bijection $\Z \rightarrow \mathbb{Z}_2^2$ defined by 
\begin{align*}
    0 \rightarrow 00, \\
    1 \rightarrow 01, \\
    2 \rightarrow 11, \\
    3 \rightarrow 10.
\end{align*}
The image $\phi(C)$ of a linear code $C$ over $\Z$ of length $n$ by the Gray map is a binary code of length $2n$. We know that $\phi$ is not only a weight-preserving map from
\begin{align*}
    (\Z^n, \text{Lee weight } (w_L)) \text{ to } (\mathbb{Z}_2^{2n}, \text{Hamming weight } (w_H))
\end{align*}
but also a distance-preserving map from
\begin{align*}
    (\Z^n, \text{Lee distance } (d_L)) \text{ to } (\mathbb{Z}_2^{2n}, \text{Hamming distance } (d_H)).
\end{align*}
We also know that for any linear code $C$ over $\Z$, its Gray map image $\phi(C)$ is distance invariant, so its minimum Hamming distance $d_H$ is equal to the minimum Hamming weight $w_H$. Thus, we know that the minimum Hamming distance of $\phi(C)$ is equal to the minimum Lee weight $d_L$ of $C$. Additionally, we know that
\begin{align*}
    w_L = w_H = d_H = d_L.
\end{align*}
Thus, $\phi(C)$ is a binary code, not necessarily linear (because the Gray map is not linear), over $\mathbb{Z}_2$ with the parameters  $[2n,2k_1+k_2,d_L]_2$. We refer the reader to \cite{Z4Book} for more details on codes over $\Z$.

\medskip

\section{Cyclic Search over $\Z$}

With the usual identification of vectors $v=(v_0,v_1,\dots,v_{n-1})$ with the corresponding polynomials $v(x)=v_0+v_1x+\cdots+v_{n-1}x^{n-1}$, cyclic codes of length $n$ over $\Z$ are ideals of the quotient ring $\Z[x]/\langle x^n-1 \rangle$. The factorization of $x^n-1$ and the algebraic structure  of $\Z[x]/\langle x^n-1 \rangle$ are similar to the field case when $n$ is odd. This is why we start the search process on cyclic codes for odd lengths.

The $\Z$ database contains data on codes for length up to 128. We first exhaustively search all cyclic codes over $\Z$ of odd lengths. Hence, we examine all cyclic codes of odd length up to 127.  To do so, we calculate all divisors of $x^n-1$ from $n=1$ to $n=127$, generate all cyclic codes, and compute their minimum Lee distances. For large lengths and dimensions, the minimum distance calculation may take too long. Therefore, we used the following restrictions.

\begin{enumerate}
\item When $n \leq 61$, we calculate Lee weight without time limit.
\item When $n \geq 63$ and $\min(2k_1+k_2,2n-2k_1-k_2) \leq 60$, we calculate Lee weight with a prescribed fixed time limit.
\item When $n \geq 63$ and $\min(2k_1+k_2,2n-2k_1-k_2) > 60$, we skip the code without calculating its Lee weight.
\end{enumerate}

The reason why our searches start with codes having odd lengths is because we have the unique factorization of $x^n-1$ for an odd $n$ and many other similarities to the field case, hence computations are easier. Specifically, we have the following theorem. 
\begin{theorem} \label{divisors} \cite{Z4QC2002} Let $n$ be an odd positive integer. Then $x^n-1$ can be factored into a product of finitely many pairwise coprime basic irreducible polynomials over $\Z$, say, $x^n-1=g_1(x)g_2(x)\cdots g_r(x)$. Also, this factorization is unique up to ordering of the factors. In fact, we have the following: if $f_2(x)|(x^n-1)$ in $\mathbb{Z}_2[x]$ then there is a unique monic polynomial $f(x) \in \Z[x]$ such that $f(x)|(x^n-1)$ in $\Z[x]$ and $f(x) = f_2(x)$, where $f_2(x)$ denotes the reduction of $f(x) \mod 2$.
\end{theorem}

\noindent The polynomial $f(x)$ in the above theorem is called the Hensel lift of $f_2(x)$. 

Other fundamental facts about cyclic codes over $\Z$ are given in the following theorems.

\begin{theorem}  \label{generator} \cite{Z4QC2002}\cite{Z4Book}\cite{CyclicZ4} Let $n$ be an odd integer and let $I$ be an ideal in $\Z[x]/\langle x^n-1 \rangle$. Then there are unique monic polynomials $f(x),g(x),h(x)$ over $\Z[x]$ such that $I=\langle f(x)h(x),2f(x)g(x) \rangle=\langle f(x)h(x)+2f(x)\rangle$  where $f(x)g(x)h(x)=x^n-1$ and $|I|=4^{\deg(g(x))}2^{\deg(h(x))}$
\end{theorem}

\begin{theorem} \cite{Z4QC2002} Let $I$ be a cyclic code over $\Z$ of odd length $n$. Then $I$ is a free module of rank $k$ if and only if the generator polynomial $p(x)$ of the corresponding ideal divides $x^n-1$ and $\deg(p(x)) = n - k$.
\end{theorem}

\noindent From the theorems above, we get the following important consequences.

\begin{corollary} Let $C$ be a cyclic code over $\Z$ of odd length $n$. Then 

\begin{enumerate}
    \item $C$ is a principal ideal.
    
    \item The number cyclic codes of length $n$ is $3^r$ and the number of free cyclic codes of length $n$ is $2^r$, where $r$ is the number of basic irreducible divisors of $x^n-1$ as in Theorem \ref{divisors}.
    
    \item  There is a one-to-one correspondence between divisors of $x^n-1$ and free cyclic codes of length $n$ over $\Z$.
\end{enumerate}

\end{corollary}

In searching for cyclic codes over $\Z$, we first found all free cyclic codes by finding all divisors of $x^n-1$ from its factorization into basic irreducibles. This is the same process as constructing cyclic codes over a finite field. We then constructed all cyclic codes that are not free using Theorem \ref{generator}. This part is different from the field case.

\section{QC Search Method}

The second search method we implemented was adapting the ASR search algorithm to $\Z$. First introduced in \cite{ASR} (named after Aydin, Siap, and Ray-Chaudhuri) and refined and generalized in more recent works (\cite{GenASR},\cite{Twistulant},\cite{2genASR}), the ASR search algorithm produced a large number of new, record breaking codes over small finite fields from the class of quasi-cyclic (QC) and quasi-twisted codes (QT). Our implementation of the ASR search algorithm for $\Z$  yielded 2369 new linear $\Z$ codes. Before describing our method, we review some basics. 

A QC code is a generalization of a cyclic code, where a cyclic shift of a codeword by $\ell$ positions gives another codeword. Such a code is called an $\ell$-QC code, or a QC code of index 2. Algebraically a QC code of length $n=m\cdot \ell$ and index $\ell$ is an $R$-submodule of $R^{\ell}$ where $R=\Z[x]/ \langle x^m-1 \rangle$. A generator matrix of a QC code can be put into the form
\vspace{-0.4 cm}
\begin{center}
\[\begin{bmatrix}
    G_{1,1} & G_{1,2} & \dots & G_{1,\ell} \\
    G_{2,1} & G_{2,2} & \dots & G_{2,\ell} \\
    \vdots  & \vdots  & \ddots&\vdots \\
    G_{r,1} & G_{r,2} & \dots & G_{r,\ell}
\end{bmatrix}
\]
\end{center}
where each $G_{i,j}=Circ(g_{i,j})$ is a circulant  matrix defined by some polynomial $g_{i,j}(x)$.  Such a code is called an $r$-generator QC code. Most of the work on QC codes in the literature is focused on the 1-generator case. A generator matrix of a 1-generator QC code can be  put in the form 
\vspace{-0.4 cm}
\begin{center}
\[\begin{bmatrix}
    G_1   G_2  \dots  G_{\ell}
\end{bmatrix}
\]
\end{center}

The class of QC codes is known to contain many codes with  good parameters. There have been various types of search algorithms on the class of QC codes. The ASR search algorithm is one that has been shown to be particularly effective for 1-generator QC codes. It is based on the following theorem.



\begin{theorem} \label{ASR} \cite{ASR} Let $C$ be a 1-generator $\ell$-QC code over $\Fq$ of length $n = m\ell$ with a generator of the form: 

$$
(f_{1}(x)g(x), f_{2}(x)g(x), ... , f_{\ell}(x)g(x)), 
$$

\noindent where $x^{m}-1= g(x)h(x)$ and  $\gcd(h(x), f_{i}(x))=1$ for all $i=1,...,\ell$. Then $C$ is an $[n, k,  d']$ code where dim($C$)=$m-\deg(g(x))$, and $d'> \ell\cdot d$ where $d$ is the minimum distance of the cyclic code $C$ of length $m$ generated by $g(x)$. 
\end{theorem}
\subsection{Adapting ASR for free cyclic codes over $\Z$}
When we adopt the ASR search algorithm for $\Z$, we consider two cases: When the starting cyclic code, that forms the the building blocks of the components of the QC code, is free and non-free. The former case is similar to the field case. To determine whether two polynomials are relatively prime over $\Z$, we use the following lemma. 
\begin{lemma} \cite{Z4Book}
Let $f_1(x)$ and $f_2(x) \in \Z[x]$ and denote their image in $\mathbb{Z}_2[x]$ under - by $\bar{f_1}(x)$ and $\bar{f_2}(x)$, respectively. Then $f_1(x)$ and $f_2(x)$ are coprime in $\Z[x]$ if and only if $\bar{f_1}(x)$ and $\bar{f_2}(x)$ are coprime in $\mathbb{Z}_2[x]$.
\end{lemma}

We begin the search process by finding all divisors of $x^m-1$ over the binary field. For each divisor $g(x)$, we compute its Hensel lift $g_4(x)$ to $\Z$. Hence, $g_4(x)$ is a divisor of $x^m-1$ over $\Z4$ and it generates a free cyclic code. We then form a generator of a QC code as in Theorem \ref{ASR} using $g_4(x)$ as the ``seed", the common term in each block. The QC codes obtained this way are also free.



\subsection{ Adapting ASR for non-free cyclic codes over $\Z$}

Adapting ASR as free cyclic codes over $\Z$ as seed only produces QC codes that are also free. It is desirable to obtain some non-free codes as well since the database is lacking non-free codes. For this we fed the same algorithm in the previous subsection with the generators (seeds) of non-free cyclic codes we obtained in the initial part of the process. We indeed obtained many new $\Z$ codes that are not free.

\section{New Linear Codes over $\Z$}


We classify the codes with good parameters that we have found into a few types: decent, good, very good, and great codes. Decent codes are those whose Gray  images are non-linear and they have minimum Lee weight $d_L$  equal to the minimum weight of the best known binary linear code. This means that if the code parameters we found are $[n,k_1,k_2,d_L]$ and parameters of the best know binary linear code are $[2n,2k_1+k_2,d]_2$ and $d = d_L$, then this code  is a decent code.

A code is a good code if it satisfies either of the two sets of conditions. First, if the code we found has a non-linear Gray map image and its parameters $[n,k_1,k_2,d_L]$  beat the parameters of the best known binary linear code $[2n,2k_1+k_2,d]_2$ but its weight  does not exceed the best known upper ($d_u$) bound on the minimum weight of a binary linear code of length $2n$ and dimension $2k_1+k_2$, then this code is a good code. Second, a code can also be a good code if it has a linear Gray map image and its parameters are the same as the parameters of the best known linear code over $\mathbb{Z}_2$.

\begin{center}
\footnotesize
\begin{longtable}{|l|l|l|l|}
\caption{Code classification via the Gray map image of code parameter $[n,k_1,k_2,d_L]$}  \label{tab:long} \\
  \hline
  \diagbox{\textbf{linearity of image}}{\boldmath{$d_L$} \textbf{from} \boldmath{$[2n,2k_1+k_2,d_L]_2$}} & \boldmath{$= d$} & \boldmath{$= d$} \textbf{but} \boldmath{$\leq d_u$} & \boldmath{$> d_u$} \\ 
  \hline
  linear & good & great & not possible \\ 
  \hline
  non-linear & decent & good & very good \\ 
  \hline
\end{longtable}
\end{center}

To be a very good code, a code needs to have non-linear Gray map image and its minimum Lee weight $d_L$ must beat the best known upper bound on the minimum weight, $d_u$, of a binary linear code over $\mathbb{Z}_2$ with the comparable  length and dimension. And to be a great code, a code needs to have not only linear Gray map image but also  parameters that beat the parameters of the best known binary linear code.

\noindent Now we give details of how many new codes we obtained from each category.

\begin{enumerate}
    \item Free cyclic codes of odd length: A total of 615 new linear $\Z$ codes of this type have been found of which 135 are decent codes, 70 good codes, and 4 very good codes. Table 2 below shows a few examples of codes from this category. All of the new codes from this category have been added to the database. We represent a polynomial as a string of its coefficients in ascending order of its terms. For example, the string 323001 on the first row of Table 2  represents the polynomial $g(x)=x^5+3x^2+2x+3$.

\begin{center}
\footnotesize
\begin{longtable}{|l|l|l|l|}
\caption{Examples of new free cyclic codes from exhaustive search} \label{tab:long} \\

\hline \multicolumn{1}{|l|}{\boldmath{$[n,k_1,k_2,d]$}} & \multicolumn{1}{l|}{\textbf{Gray map image}} & \multicolumn{1}{l|}{\textbf{comparison with BKLC}} & \multicolumn{1}{l|}{\boldmath{$g$}} \\ \hline
\endfirsthead


\hline \multicolumn{4}{|r|}{{Continued on next page}} \\ \hline
\endfoot

\hline
\endlastfoot

$[31,26,0,4]$ & non-linear & decent code & 323001\\
\hline
$[47,24,0,16]$ & non-linear & good code & 331123310332331020110201\\
\hline
$[117,90,0,6]$ & non-linear & — & 3020330000100110222210012321\\

\end{longtable}
\end{center}
    
    \item Non-free cyclic codes of odd length:  A total of 5717 new linear $\Z$ codes of this type have been found of which 134 are decent codes and 271 are good codes.  Table 3 below shows a few examples of codes from this category. All of the new codes from this category have been added to the database.
    
\begin{center}
\footnotesize
\begin{longtable}{|l|l|l|l|}
\caption{Examples of new nonfree cyclic codes from exhaustive search} \label{tab:long} \\

\hline \multicolumn{1}{|l|}{\boldmath{$[n,k_1,k_2,d]$}} & \multicolumn{1}{l|}{\textbf{Gray map image}} & \multicolumn{1}{l|}{\textbf{comparison with BKLC}} & \multicolumn{1}{l|}{\boldmath{$g$}} \\ \hline
\endfirsthead


\hline \multicolumn{4}{|r|}{{Continued on next page}} \\ \hline
\endfoot

\hline
\endlastfoot

$[21,17,4,2]$ & linear & good code & 32311\\
\hline
$[45,24,1,8]$ & non-linear & — &  1201112212020113303211\\
\hline
$[105,19,1,44]$ & non-linear & — &  \makecell{320232031230302133230113333002 \\ 321201321010311333010302003000 \\ 100030131000101002301110101}\\
\hline
$[125,120,5,2]$ & linear & good code & 100001\\

\end{longtable}
\end{center}
    
    \item QC codes where $m$ is odd, and the seed codes are free cyclic codes: A total of 452 new $\Z$-linear codes of this type have been found of which  76 are decent codes, 3 are good codes, and 1 is a very good code. Table 4 below shows a few examples of codes from this category. All of the new codes from this category have been added to the database.
    
\begin{center}
\footnotesize
\begin{longtable}{|l|l|l|l|l|l|}
\caption{Examples of new free QC codes found via ASR algorithm} \label{tab:long} \\

\hline \multicolumn{1}{|l|}{\boldmath{$[n,k_1,k_2,d]$}} & \multicolumn{1}{l|}{\boldmath{$l$}} & \multicolumn{1}{l|}{\textbf{Gray map image}} & \multicolumn{1}{l|}{\textbf{comparison with BKLC}} & \multicolumn{1}{l|}{\boldmath{$g$}} & \multicolumn{1}{l|}{\boldmath{$f$}}\\ \hline
\endfirsthead


\hline \multicolumn{6}{|r|}{{Continued on next page}} \\ \hline
\endfoot

\hline
\endlastfoot

$[22,10,0,12]$ & 2 & non-linear & decent code & 31 &  \parbox{3cm}{\begin{math}\begin{aligned}
~\\[-3ex]
&f_1 = 2101311121 \\
&f_2 = 1123112011
\end{aligned}\end{math}}\\
\hline
$[30,9,0,18]$ & 2 & non-linear & — & 1021311 &  \parbox{3cm}{\begin{math}\begin{aligned}
~\\[-3ex]
&f_1 = 01030023 \\
&f_2 = 31003013
\end{aligned}\end{math}}\\
\hline
$[35,4,0,32]$ & 7 & non-linear & decent code & 31 & 
\parbox{3cm}{\begin{math}\begin{aligned}
~\\[-3ex]
&f_1 = 0303 \\
&f_2 = 3221 \\
&f_3 = 102 \\
&f_4 = 311 \\
&f_5 = 2311 \\
&f_6 = 3213 \\
&f_7 = 33
\end{aligned}\end{math}}\\
\hline
$[54,21,0,22]$ & 2 & non-linear & — & 1001001 &  \parbox{3cm}{\begin{math}\begin{aligned}
~\\[-3ex]
&f_1 = 2321012031 \\
&3033223332 \\
&2 \\
&f_2 = 2320013322 \\
&3130002020 \\
&2
\end{aligned}\end{math}}\\
\hline
$[75,10,0,54]$ & 5 & non-linear & — & 321231 & \parbox{3cm}{\begin{math}\begin{aligned}
~\\[-3ex]
&f_1 = 1230312011 \\
&f_2 = 2332233233 \\
&f_3 = 0022320232 \\
&f_4 = 1302320302 \\
&f_5 = 2113222122
\end{aligned}\end{math}}\\

\end{longtable}
\end{center}
    
    \item QC codes where the seed polynomials generate  non-free cyclic codes: A total of 1917 new $\Z$ linear codes of this type have been found of which 119 are decent codes, 127 are good codes, and 1 is a very good code. Table 5 below shows a few examples of codes from this category. All of the new codes from this category have been added to the database.

\begin{center}
\footnotesize
\begin{longtable}{|l|l|l|l|l|l|}
\caption{Examples of new non-free QC codes found via the ASR algorithm} \label{tab:long} \\

\hline \multicolumn{1}{|l|}{\boldmath{$[n,k_1,k_2,d]$}} & \multicolumn{1}{l|}{\boldmath{$l$}} & \multicolumn{1}{l|}{\textbf{Gray map image}} & \multicolumn{1}{l|}{\textbf{comparison with BKLC}} & \multicolumn{1}{l|}{\boldmath{$g$}} & \multicolumn{1}{l|}{\boldmath{$f$}}\\ \hline
\endfirsthead


\hline \multicolumn{6}{|r|}{{Continued on next page}} \\ \hline
\endfoot

\hline
\endlastfoot

$[6,1,2,4]$ & 2 & linear & — & 311 & 
\parbox{2cm}{\begin{math}\begin{aligned}
~\\[-3ex]
&f_1 = 3 \\
&f_2 = 3
\end{aligned}\end{math}}\\
\hline
$[28,0,3,32]$ & 4 & linear & good code & 31101 & 
\parbox{2cm}{\begin{math}\begin{aligned}
~\\[-3ex]
&f_1 = 2 \\
&f_2 = 222 \\
&f_3 = 202 \\
&f_4 = 022
\end{aligned}\end{math}}\\
\hline
$[45,5,0,40]$ & 3 & non-linear & decent code & \makecell{3032233011 \\ 1} & 
\parbox{2cm}{\begin{math}\begin{aligned}
~\\[-3ex]
&f_1 = 30121 \\
&f_2 = 21021 \\
&f_3 = 30103
\end{aligned}\end{math}}\\
\hline
$[63,4,9,40]$ & 3 & non-linear & — & \makecell{1323002332 \\ 10003121} & 
\parbox{2cm}{\begin{math}\begin{aligned}
~\\[-3ex]
&f_1 = 3021 \\
&f_2 = 3303 \\
&f_3 = 1211
\end{aligned}\end{math}}\\
\hline
$[66,1,12,44]$ & 2 & linear & — & \makecell{3001023221 \\ 2032230010 \\ 2100100100 \\ 1} & 
\parbox{2cm}{\begin{math}\begin{aligned}
~\\[-3ex]
&f_1 = 111 \\
&f_2 = 331
\end{aligned}\end{math}}\\
\hline
$[112,4,3,92]$ & 16 & non-linear & — & 1121 & 
\parbox{2cm}{\begin{math}\begin{aligned}
~\\[-3ex]
&f_1 = 3111 \\
&f_2 = 3332 \\
&f_3 = 1001 \\
&f_4 = 0311 \\
&f_5 = 1033 \\
&f_6 = 3011 \\
&f_7 = 0213 \\
&f_8 = 0121 \\
&f_9 = 3131 \\
&f_{10} = 0313 \\
&f_{11} = 3213 \\
&f_{12} = 1132 \\
&f_{13} = 3211 \\
&f_{14} = 1032 \\
&f_{15} = 1101 \\
&f_{16} = 0113
\end{aligned}\end{math}}\\

\end{longtable}
\end{center}

    \item New binary non-linear codes: We found 2 new binary non-linear codes based on the comparison with the database \cite{NonlinearBinaryTable}. These two codes however  have the same parameters as the corresponding BKLCs. Apparently, the database \cite{NonlinearBinaryTable} has not been updated for a long time and it does not seem to incorporate the data from \cite{database}. The database \cite{NonlinearBinaryTable} was constructed to give information on $A(n,d)$, the size of the largest binary code of length $n$ and minimum distance $d$. As such, both linear and non-linear codes must be considered in determining $A(n,d)$. Hence, the data from \cite{database} must be taken into account in determining $A(n,d)$. Of the two codes we have found, one  is from the exhaustive search of non-free cyclic codes.

\begin{center}
\footnotesize
\begin{longtable}{|l|l|l|l|}
\caption{The new non-linear binary code from exhaustive searches of non-free cyclic codes} \label{tab:long} \\

\hline \multicolumn{1}{|l|}{\boldmath{$[n,k_1,k_2,d]$}} & \multicolumn{1}{l|}{\textbf{Gray map image}} & \multicolumn{1}{l|}{\textbf{comparison with BKLC}} & \multicolumn{1}{l|}{\boldmath{$g$}}\\ \hline
\endfirsthead


\hline \multicolumn{4}{|r|}{{Continued on next page}} \\ \hline
\endfoot

\hline
\endlastfoot

$[51,10,8,28]$ & non-linear & decent code & \makecell{10000012131032001222 \\ 23001111010223122032 \\ 31} \\

\end{longtable}
\end{center}
    
  \noindent  The other code is from a search of free QC codes by the ASR algorithm.
    
\begin{center}
\footnotesize
\begin{longtable}{|l|l|l|l|l|l|}
\caption{New non-linear binary code from search of free QC codes} \label{tab:long} \\

\hline \multicolumn{1}{|l|}{\boldmath{$[n,k_1,k_2,d]$}} & \multicolumn{1}{l|}{\boldmath{$l$}} & \multicolumn{1}{l|}{\textbf{Gray map image}} & \multicolumn{1}{l|}{\textbf{comparison with BKLC}} & \multicolumn{1}{l|}{\boldmath{$g$}} & \multicolumn{1}{l|}{\boldmath{$f$}}\\ \hline
\endfirsthead


\hline \multicolumn{6}{|r|}{{Continued on next page}} \\ \hline
\endfoot

\hline
\endlastfoot

$[51,16,0,26]$ & 3 & non-linear & decent code & 31 & 
\parbox{4cm}{\begin{math}\begin{aligned}
~\\[-3ex]
&f_1 = 3223033120003033 \\
&f_2 = 2122003313031103 \\
&f_3 = 0232111300112321
\end{aligned}\end{math}}\\

\end{longtable}
\end{center}
   
\end{enumerate}

\section{Conclusion and Further Research}
Codes over $\Z$ have a special place in coding theory. A database of $\Z$ codes was constructed in 2008 but it was moved to a new server with a new URL recently. We have updated the original database of $\Z$ codes substantially by adding 8701 new codes.  Of these, a total 433 are decent codes, 426 are good codes,  5 are very good codes, and 2 are new non-linear binary codes. These codes have been found by implementing search algorithms for cyclic and QC codes. Still more codes can be found and added to the database by conducting similar searches by considering cyclic codes of even length and related searches for QC. We also note that the database \cite{NonlinearBinaryTable} of non-linear binary codes is in need of an update.



\end{document}